# Insights from early mathematical models of 2019-nCoV acute respiratory disease (COVID-19) dynamics

Jomar F. Rabajante[1,2,3,*]

Professor 2

[1]Biomathematics Research Cluster, Institute of Mathematical Sciences and Physics, University of the Philippines Los Baños, 4031 Laguna Philippines

[2]Computational Interdisciplinary Research Labs, University of the Philippines Los Baños, 4031 Laguna Philippines

[3]Faculty of Education, University of the Philippines Open University, 4031 Laguna Philippines

*jfrabajante@up.edu.ph

**Insights from early mathematical models of 2019-nCoV acute respiratory disease (COVID-19) dynamics**

## ABSTRACT

In December 2019, a novel coronavirus (SARS-CoV-2) has been identified to cause acute respiratory disease in humans. An outbreak of this disease has been reported in mainland China with the city of Wuhan as the recognized epicenter. The disease has also been exported to other countries, including the Philippines, but the level of spread is still under control (as of 08 February 2020). To describe and predict the dynamics of the disease, several preliminary mathematical models are formulated by various international study groups. Here, the insights that can be drawn from these models are discussed, especially as inputs for designing strategies to control the epidemics. Proposed model-based strategies on how to prevent the spread of the disease in local setting, such as during large social gatherings, are also presented. The model shows that the exposure time is a significant factor in spreading the disease. With a basic reproduction number equal to 2, and 14-day infectious period, an infected person staying more than 9 hours in the event could infect other people. Assuming the exposure time is 18 hours, the model recommends that attendees of the social gathering should have a protection with more than 70 percent effectiveness.

**INTRODUCTION**

A new coronavirus has been identified to cause respiratory illness, such as an atypical pneumonia, in humans (*European Centre for Disease Prevention and Control 2020*). This disease, with the interim name '2019-nCoV acute respiratory disease (ARD)' [official name: COVID-19], is first detected in the winter month of December 2019 in a city of 11 million people – Wuhan in Hubei Province, China (*World Health Organization 2020a*, *Tweeten et al. 2020*). The 2019-nCoV ARD is believed to be zoonotic in origin, from bats to intermediate host to humans (*Zhou P. et al. 2020*); and its initiation is geographically associated, but with uncertainty, with the Huanan Seafood Market in Wuhan (*Cohen 2020a*). Human-to-human transmission of 2019-nCoV has been established, such as through respiratory droplets (*Chan et al. 2020*), and there is also a suspicion of asymptomatic infection (*Chan et al. 2020*, *Kupferschmidt 2020*). To contain the epidemics, the government of China has ordered cancellation of huge events for the Chinese New Year celebration, and the lockdown of Wuhan and other cities (*Tweeten et al. 2020*).

The disease has been exported to other parts of China and to other countries (Singapore, Thailand, Japan, South Korea, Australia, Germany, USA, Philippines etc.) generally via travel-related activities (*Johns Hopkins CSSE 2020*, *Gardner 2020*). The World Health Organization (WHO) has declared the 2019-nCoV ARD outbreak as a 'Public Health Emergency of International Concern' on 30 January 2020, specifically to enhance the level of preparedness of countries that need additional support (*World Health Organization 2020b*). To prevent the global spread of the virus, many countries have imposed travel restrictions to and from China (*Whitley et al. 2020*). The Philippine government issued temporary travel ban starting on 02 February 2020 to flights coming from and going to China, including Hong Kong and Macau (*Whitley et al. 2020*, *ABS-CBN 2020*). During the travel ban, foreign nationals who have been to China within 14 days prior to their arrival to the Philippines will be denied entry; but Filipino citizens and those who have a permanent resident visa may enter Philippines, but they need to undergo a 14-day quarantine (*ABS-CBN 2020*).

The World Health Organization has requested all governments and public health institutions to "implement decisions that are evidence-based" (*World Health Organization 2020b*). However, there are limited data and information on 2019-nCoV ARD, especially during the initial stage of its epidemics where antiviral treatment and vaccine are not yet available. To contribute in addressing the challenge of predicting the spread of novel diseases, and assessing the possible risks and interventions, mathematical models and simulation apps can be used (*Siettos and Russo 2013*, *Heesterbeek et al. 2015*, *Tuite and Fisman 2020*). Based on the results of the models, it is possible to prescribe possible epidemiological and social strategies to effectively and efficiently prevent or control emerging diseases that could initiate pandemics (*Heesterbeek et al. 2015*). Mathematical modelers can be valuable non-frontline members of the 'early responders' team during an outbreak.

## MATERIALS AND METHODS

Selected preliminary models of 2019-nCoV ARD, published in official webpages of academic/research institutions, as preprints, or as journal articles on or before 08 February 2020 (with updates on 11 February 2020), are reviewed. Important insights from the results of these models are discussed.

Moreover, a model using a Susceptible-Exposed-Infected (SEI) framework is formulated to propose measures to prevent epidemics during large events, e.g., during parties or concerts with huge crowds. The parameter values used in this model are based on the known information about 2019-nCoV ARD. The model is represented by a system of differential equations, and simulated using Berkley Madonna for Mac version 9.1.19.

## RESULTS AND DISCUSSIONS

Various international study groups have attempted to model the dynamics of the 2019-nCoV ARD. It is important to note that it might still be early to predict what will happen due to the complexity of the biomedical factors and social systems involved. Mathematical modelers are not fortune tellers, and in fact, modelers do not wish their worst-case prediction to happen. Models can be used as intelligent input during decision-making. Here are the preliminary models of 2019-nCoV ARD published on or before 11 February 2020.

### A. Model by *Gardner (2020)*

The team of Lauren Gardner of the Center for Systems Science and Engineering, Johns Hopkins University used a stochastic simulation model which is intended to mitigate pandemics at the onset of the outbreak (*Gardner 2020*, *Zlojutro et al. 2019*). Their metapopulation model connects airport networks (connections are weighted based on travel volume) at global scale. In each airport, a discrete-time Susceptible-Exposed-Infected-Recovered (SEIR) model is implemented. To model the 2019-nCoV ARD spread, the parameters and assumptions used are Assumption i. incubation period: 5 days;

Assumption ii. effective contact rate (related to the basic reproductive number $R_0$): 2;

Assumption iii. recovery period: 5 days; and

Assumption iv. initial cases are only from Wuhan with no border control.

*Gardner (2020)* estimated the average number (based on 250 runs) of infecteds in mainland China and the distribution of infected travelers worldwide. Their estimates are as follow:

a. Predicted cumulative number of infecteds in China by 31 January 2020 is 58,000 (actual reported number is 12,000). They attributed part of the discrepancy to reporting delays, suggesting that there can be mild or asymptomatic cases who do not seek medical care and are not reported.
b. The predicted start of the epidemics happened in November 2019, with hundreds of infecteds already present in early December.
c. Thailand, Taiwan, Hong Kong, South Korea, Singapore, Japan, Macau, USA, Australia and France are the top 10 countries with most number of predicted imported cases by 31 January 2020. The predicted number of imported cases in Thailand is approximately 25.

*Gardner (2020)* concluded that "the actual number of 2019-nCoV cases in mainland China are likely much higher than that reported to date." The outbreak control assumed is through passenger screening upon arrival at airports. The team from Center for Systems Science and Engineering, Johns Hopkins University also developed an interactive dashboard that map the 2019-nCoV ARD outbreak in almost real-time mode (*Johns Hopkins CSSE 2020*).

B. Model by *Imai et al. (2020a, 2020b)*

The team of Natsuko Imai and Neil M. Ferguson from WHO Collaborating Centre for Infectious Disease Modelling, MRC Centre for Global Infectious Disease Analysis, Imperial College London published several sets of reports. One is on the transmissibility of 2019-nCoV (*Imai et al. 2020a*), and another on estimating the potential total number of 2019-nCoV ARD cases in Wuhan (*Imai et al. 2020b*).

In the first study (*Imai et al. 2020a*), the conclusions are as follow:

a. Estimated basic reproductive number $R_0$ is 2.6 with uncertainty range of 1.5-3.5 (as of 18 January 2020). $R_0$ is the average number of people that one case can infect in a population full of susceptibles, excluding the secondary cases. An $R_0 > 1$ indicates potential outbreak.
b. It is suggested to block >60 percent of transmission to effectively control the outbreak. This conclusion is probably because the rule of thumb for calculating critical elimination threshold in a well-mixed host population (minimum fraction of the susceptibles that needs to be successfully protected, e.g., by vaccination to attain herd immunity) is

   $p_c = 1 - {}^{1}/_{R_0}$ (*Heesterbeek et al. 2015*).

In the second study (*Imai et al. 2020b*), they concluded that the estimated total number of symptomatic infecteds in Wuhan as of 18 January 2020 is 4,000 people with uncertainty range of 1,000-9,700. They used the following assumptions:

Assumption i. Wuhan International Airport serves 19 million individuals (catchment size) from the city of Wuhan and neighboring locations;

Assumption ii. delay between infection to detection: 10 days;

Assumption iii. incubation period: 5-6 days;

Assumption iv. international travels from Wuhan: 3,301 passengers per day;

Assumption v. exit screening implemented on 15 January 2020 has no impact on exported cases as of 16 January 2020; and

Assumption vi. infecteds travelling outside mainland China are detected at their destinations.

C. Other models on estimating the basic reproductive number and its implications (*Zhao et al. 2020a*, *Shen et al. 2020*, *Li et al. 2020*, *Wu et al. 2020*, *Riou and Althaus 2020*, *Read et al. 2020*, *Zhou T. et al. 2020*, *Kucharski et al. 2020*, *Park 2020*)

There are multiple models that aim to estimate the basic reproduction number $R_0$ of 2019-nCoV ARD even at the early phase of the outbreak. Shi Zhao from JC School of Public Health and Primary Care, Chinese University of Hong Kong and colleagues estimated $R_0$ using the exponential growth model (*Zhao et al. 2020a*). The model assumed serial interval for 2019-nCoV to be 8.0$\pm$3.6 days, which is estimated based on the serial intervals of MERS-CoV and SARS. Serial interval is the duration between the onset of symptoms of the primary case and the onset of symptoms of the secondary case infected by the primary. The results of the study are (*Zhao et al. 2020a*):

a. The estimated mean $R_0$ ranges from 2.24 to 5.71 based on the reporting rate of cases.
b. If there is no change in the reporting rate (as of 10-24 January 2020), the estimated $R_0$ is 5.71, which is estimated to be between MERS-CoV-like (5.31) and SARS-like (6.11) respiratory syndrome.
c. If the reporting rate will increase 2-fold, the estimated $R_0$ is 3.58, which is estimated to be between MERS-CoV-like (3.38) and SARS-like (3.77) respiratory syndrome.
d. If the reporting rate will increase 8-fold, the estimated $R_0$ is 2.24, which is estimated to be between MERS-CoV-like (2.16) and SARS-like (2.32) respiratory syndrome.

A comparison of the pathogenicity and transmissibility of 2019-nCoV with other viruses can be found in the paper by *Chen (2020)*. Based on clinical studies, 2019-nCoV is less pathogenic than MERS-CoV and SARS, but there is still debate on the transmissibility of 2019-nCoV with estimated $R_0$ ranging 1.4-5.5 (SARS is 2-5 and MERS-CoV is <1) (*Chen 2020*). The estimated case fatality rate for 2019-nCoV is 3 percent (10% for SARS and 40% for MERS-CoV) (*Chen 2020*). Case fatality rate is the number of deaths due to the disease divided by the total number of cases having the disease. It should be noted that case fatality rate is not equal to an infected's chance of death due to the disease (although many use this as approximation). Probability of death or survival is person or situation-dependent. Also, a case fatality rate computed at global or national scale could be different from the case fatality rate computed at local scale. If 2019-nCoV ARD outbreak happens in a community with inferior health system, then case fatality rate in that community may be higher. The case fatality rate currently being computed is still an initial estimate since the epidemics is still ongoing, the number of deaths and total number of cases could still change.

*Shen et al. (2020)* predict epidemic of 2019-nCoV in China totalling to 8,042 infecteds with case fatality rate of 11.02 percent. Their prediction used data prior to 25 January 2020, and is not anymore accurate as of 8 February 2020. As of 08 February 2020, total confirmed cases in mainland China is 34,611, total deaths is 723 and total recovered is 2,370 (*Johns Hopkins CSSE 2020*). *Shen et al. (2020)* also estimated a basic reproduction number $R_0$ = 4.71 at the start of the epidemic on 12 December 2019 and the effective reproduction number $R_e$ decreasing to 2.08 as of 22 January 2020. $R_e$ is the average number of new cases caused by an infected when only a fraction of the population is susceptible.

*Shen et al. (2020)* predict that if effective intervention continues, the epidemics is expected to peak in March 2020. They suggested that every one day reduction in the average duration from disease onset to isolation of infecteds will reduce the peak population size by 72-84 percent, and both the cumulative infected cases and deaths by 68-80 percent. Also, every 10 percent reduction in transmission rate could reduce the peak population size by 20-47 percent, and both the cumulative infected cases and deaths by 23-49 percent.

*Li et al. (2020)* estimated a mean incubation period of 5.2 days for 2019-nCoV ARD, mean serial interval of 7.5 days, and $R_0$ = 2.2. They also predict that cases in Wuhan is doubling in size approximately every 7.4 days.

Joseph T. Wu, Kathy Leung and Gabriel M. Leung of WHO Collaborating Centre for Infectious Disease Epidemiology and Control, School of Public Health, Li Ka Shing Faculty of Medicine, University of Hong Kong (*Wu et al. 2020*) presented their 'nowcast' and forecast of the 2019-nCoV ARD outbreak. They used a Susceptible-Exposed-Infectious-Recovered (SEIR) metapopulation model, considering the metropolitan-wide quarantine of Wuhan and neighboring cities, the international air travels to and from China, and the human mobility in mainland China. Using Markov Chain Monte Carlo method, the estimated posterior mean of $R_0$ = 2.68 (with 95% credible interval 2.47-2.86). The model predicts that there are 75,815 infecteds in Wuhan as of 25 January 2020 which is way larger than the number of reported cases, and the epidemic doubling time is 6.4 days. They also predict that, if transmissibility in cities in mainland China is similar to Wuhan, then localized outbreaks are being sustained and the epidemics is already growing exponentially with a lag time of 1-2 weeks behind the Wuhan outbreak.

*Wu et al. (2020)* predict that large cities with close transportation links with China may become another epicenter of the epidemics unless effective control strategies are placed. Without reduction in mobility and transmissibility, daily 2019-nCoV ARD incidence will peak in the last week of March and early week of April 2020 with around 35 cases per 1,000 population. Reducing transmissibility by 25 percent will reduce the daily incidence to around 20 cases per 1,000 population and delay the peak of the outbreak to early May 2020.

Julien Riou and Christian L. Althaus of the Institute of Social and Preventive Medicine, University of Bern predict that $R_0$ is around 2.2 (90% high density interval 1.4–3.8) (*Riou and Althaus 2020*). *Riou and Althaus (2020)* also estimated the median dispersion parameter $k$ = 0.54 that measure the chance of super-spreading ($k$ is a parameter of the negative-binomial offspring distribution where value near zero denotes overdispersion). They indicated that transmission characteristics of 2019-nCoV is similar to SARS and to the 1918 pandemic influenza. They suggested heightened screening, surveillance and control at airports and other travel hubs to prevent further global outbreak.

*Read et al. (2020)* estimated an $R_0$ between 3.6 and 4.0, with corresponding critical elimination threshold $p_c$ between 72-75 percent. This means that containment of the virus could be challenging. They estimated that only 5.1 percent of infecteds in Wuhan are reported, and by 4 February 2020, there are >190k cases in Wuhan. Major cities like Shanghai, Beijing, Guangzhou, Chongqing and Chengdu are predicted to have large outbreaks. Thailand, Japan, Taiwan, Hong Kong and South Korea have high risk of disease importation via air travel. The lockdown of Wuhan is estimated to contribute only to 24.9 percent reduction in epidemic size in locations outside Wuhan.

*Zhou T. et al. (2020)* used SEIR to estimate $R_0$ which is in the range of 2.2 and 3.0. Moreover, the group of Adam J. Kucharski also estimated median $R_0$ fluctuating between 1.6-2.9 (*Kucharski et al. 2020*). *Kucharski et al. 2020* used an SEIR-based stochastic transmission model to determine transmission over time, prevalence of symptomatic cases, and importation of the disease as of 23 January 2020. The assumptions of the model by *Kucharski et al. 2020* are as follow:

- Assumption i. incubation period is Erlang distributed (rate=2) with mean of 5.2 days;
- Assumption ii. delay from symptom onset to isolation follows an Erlang distribution (rate=2) with mean of 2.9 days;

Assumption iii. delay from symptom onset to reporting follows an exponential distribution with mean 6.1 days;

Assumption iv. proportion of reported cases = 16 percent;

Assumption v. probability of disease being exported from Wuhan to a certain location depends on the number of cases in Wuhan, the air travel connections and volume between the two places (3,300 passengers per day) before the lockdown on 23 January 2020, and the probability of reporting an imported case;

Assumption vi. transmission is modeled as geometric random walk process as

$$d \log \beta = \alpha \, dB_t$$

where $\beta$ is the transmission rate in the force of infection, $\alpha$ is the volatility and $B_t$ is Brownian motion;

Assumption vii. sequential Monte Carlo is used to estimate the evolving transmission rate and daily effective reproduction number;

Assumption viii. the outbreak is initiated by a single case on 02 December 2019 (sensitivity analysis is also done by changing to 10 initial cases);

Assumption ix. super-spreading is modeled using a branching process with negative-binomial offspring distribution. The probability that an outbreak will occur after *n* introductions is computed as

$$1 - (1 - P_e)^n$$

where $P_e$ is the probability that a single case will initiate an outbreak;

Assumption x. latent period is equal to the incubation period (which means asymptomatics are non-infectious); and Assumption xi. the model has two sets of compartments: one for the population in Wuhan, and one for international travelers. The compartment for the population in Wuhan has SEIR sub-compartments; while EIR sub-compartments for the international travelers.

Combining multiple sources of data, the results of *Kucharski et al. (2020)*, in addition to the $R_0$ estimate, are as follow:

a. their model replicated the reported temporal trend of epidemics in Wuhan and other countries as of 30 January 2020;
b. if there are >3 cases in a location with similar transmissibility as in Wuhan, then there is >50 percent probability of outbreak in that location; and
c. there will be substantial variation in the transmissibility of 2019-nCoV over time, which suggests that effectiveness of control measures is setting-dependent.

Due to the uncertainties in the early stages of an epidemic that could lead to pandemic, estimating transmissibility and severity (e.g., by analyzing data of the first few hundred cases) can be a challenging task. *Black et al. (2017)* proposed a way to do an early estimation. This could help authorities in estimating the impact and level of risks. To reconcile the different $R_0$ estimates for the 2019-nCoV ARD epidemics, *Park et al. (2020)* proposed a framework to compare and integrate the different $R_0$ values estimated using different models.

D. Models by *De Salazar et al. (2020)* and *Chinazzi et al. (2020)*

The team of Pablo M. De Salazar and Rene Niehus from Center for Communicable Disease Dynamics, Department of Epidemiology, Harvard T.H. Chan School of Public Health used simple regression analysis to predict imports of 2019-nCoV cases from Wuhan to other countries outside China *(De Salazar et al. 2020)*. The goal of their study is to determine possible countries that are under-detecting the number of imported cases. The assumptions of their model are:

Assumption i. only direct air travel volume from Wuhan to other countries are considered; and

Assumption ii. the observed case counts follow a Poisson distribution. This means that *De Salazar et al. (2020)* assumes that the case counts are random and have equal mean and variance.

*De Salazar et al. (2020)* used generalized linear regression, which may only detect linear but not the nonlinear relationship in the data. The regression model with coefficient $\beta$ is as follows:

$$C_i \sim Poisson(\lambda_i)$$

$$\lambda_i = \beta(x_i + 1)$$

where $C_i$ is the observed case count in the i-th country which is assumed to follow a Poisson distribution with estimated mean $\lambda_i$. The variable $x_i$ is the number of daily flight passengers from Wuhan to the i-th country. Inferring from the results of *De Salazar et al. (2020)*, the maximum likelihood estimate of $\beta$ is 0.07 with bounds 0.04-0.13 (95% prediction interval). Based on the model by *De Salazar et al. (2020)*, Indonesia and Cambodia as well as Thailand may have undetected cases.

Other models have also considered importation of 2019-nCoV ARD from Wuhan to other countries using transportation data (*Gardner 2020, Yuan et al. 2020, Chinazzi et al. 2020*). *Chinazzi et al. (2020)* ranked cities based on associated risk of disease importation, with Shanghai (risk value = 5.7%) and Beijing (risk value = 5.1%) as the top two in mainland China; and Hong Kong (risk value = 6.6%), Bangkok (risk value = 1.4%), Seoul (risk value = 0.6%), Taipei (risk value = 0.6%) and Tokyo (risk value = 0.5%) as the top 5 cities outside mainland China. Assuming that 20 million passengers is the catchment size of Wuhan International Airport, the estimated median outbreak size is 31,200 cases (95% credible interval 23,400-40,400).

E. Model by *Nishiura et al. (2020)*

It is important to know the proportion of reported cases vs total number of cases (*Nishiura et al. 2020*, *Zhao et al. 2020b*). *Nishiura et al. (2020)* estimated the ascertainment rate of infection in Wuhan at 9.2 percent (95% confidence interval 5.0-20.0). This rate informs us that 90 percent of the cases is potentially undiagnosed or unreported. The infection fatality risk is estimated at 0.3-0.6 percent, comparable to 1957-1958 Asian influenza pandemic (*Nishiura et al. 2020*, *Jung et al. 2020*).

In the model by *Nishiura et al. (2020)*, they used the data from the 565 Japanese passengers evacuated from Wuhan who were screened for symptoms using portable thermo-scanners. There were 63 who showed symptoms. The passengers were also tested for the presence of 2019-nCoV using reverse transcription polymerase chain reaction (RT-PCR). Eight were positive with five of them are asymptomatic. Assuming the population size in Wuhan is 11 million, the balance equation for the risk of infection used by *Nishiura et al. (2020)* is

$$\frac{8}{565} = \frac{c(t) \times T}{q \times 11\, million}$$

where $c(t)$ is the cumulative number of reported cases in Wuhan as of time $t$ (as of 29 January 2020, there were 1,905 confirmed cases), $T$ is the window of virus detection based on the serial interval (e.g., $T$ = 7.5 days), and $q$ is the ascertainment rate. The cumulative incidence can be estimated by ${c(t)}/{q}$. Due to the limitations of diagnosis techniques, small $T$ (short detection window) may result in low value of $q$.

F. Model by *Ming et al. (2020)*

The model by *Ming et al. (2020)* aimed at forecasting the actual number of infecteds in order to estimate the number of beds needed in isolation wards and intensive care units (ICUs) in Wuhan. They used a Susceptible-Infected-Recovered model. The results of their model are as follow:

a. Assuming 50 percent detection rate without public health interventions, the actual number of cases could be greater than the number of reported cases. The forecast is 88,075 cases as of 31 January 2020 where 34,786 cases in isolation wards and 9,346 in ICUs.
b. If public health interventions are implemented with 70 percent efficacy, the forecast can be reduced (26,498 total cases as of 31 January 2020). The implication of this is that continued large-scale anti-transmission controls should be executed across all members of the population (e.g., closure of schools and suspension of public transport).

An outbreak may disrupt the activities of a community. From the models discussed above, it can be observed how mathematical modelers can help in crafting decisions or policies to prevent or control infectious diseases. In the next model to be presented, the author formulated a model that can be used to recommend auxiliary strategies to prevent the spread of 2019-nCoV ARD in large social gatherings, especially during the early period of a possible outbreak. The following model can be used to inform the public on possible scenarios that may happen during social events, and the quantitative insights drawn from the model results can be translated to actual strategies.

G. A model for preventing disease spread in large social gatherings

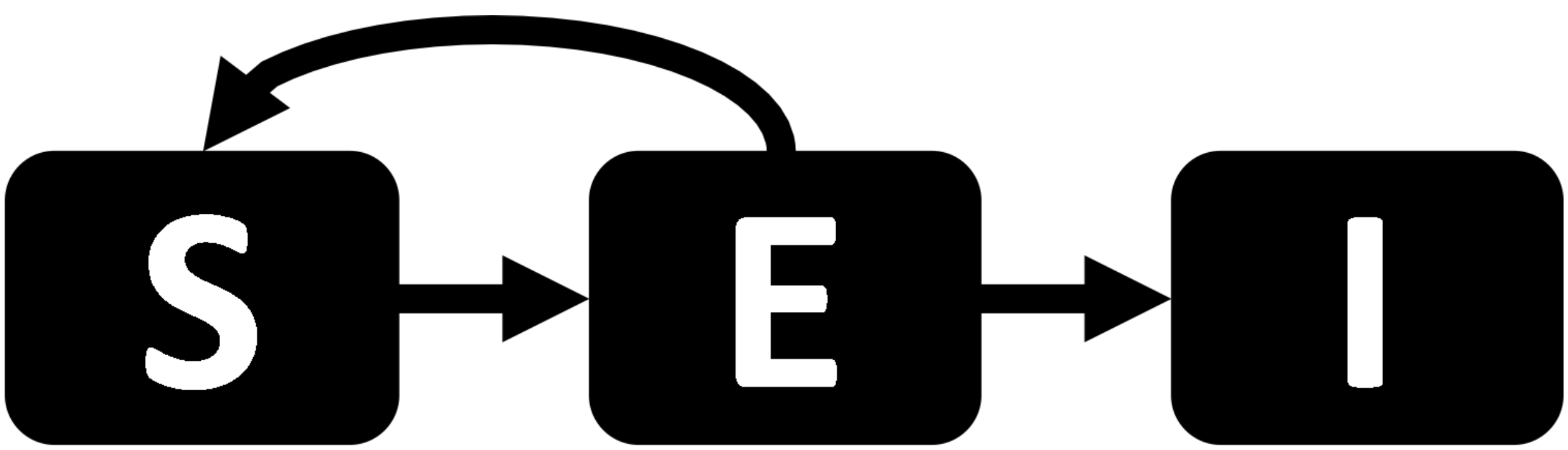

**Figure 1**. Susceptible-Exposed-Infected model framework. An infected may shed virus during the social gathering and a susceptible may be exposed. An exposed individual can be protected from being infected by having a protection (e.g., by washing hands or using face mask)

In this model, a simple Susceptible-Exposed-Infected (SEI) framework (Figure 1) is used to propose measures to prevent epidemics during large events, e.g., during parties or concerts with huge crowds. The aim of the model is prevention so the Recovered compartment in SEIR framework is not considered, hence the simulation runs do not predict the whole disease outbreak in a community. The focus of the model is on disease transmission during an event that is short in duration (max 24 hours) in a population that is formed randomly (well-mixed).

Let us suppose $\alpha$ is the level of protection of exposed persons. For simplicity, the net arrival-departure rate of attendees is denoted by integer $c$. The $\beta$ in the force of infection is derived based on the known $R_0$ (e.g., $R_0$ = 2) and population size of the susceptibles ($S_0$) as follows

$$\beta = \frac{R_0}{\tau} + \frac{R_0}{\tau S_0}.$$

The $\tau$ is a timescale and tuning factor to adjust the parameters to the unit of time used (i.e., 'hours'). The assumed default value for $\tau$ is $14 \times 24 = 336$, assuming $R_0$ is computed for a 14-day infectious period with 1 day = 24 hours. If the assumed infectious period associated with $R_0$ is decreased, the disease transmission rate in the model will increase (user of this model can vary the value of $\tau$). The model also includes possible effect of super-spreading of select individuals in the $R_0$; the new expression is $R = 2 + Poisson(0.25)$, where $Poisson(0.25)$ is a Poisson distributed random number with assumed mean of 0.25. The exposure time or the number of hours an infectious person is staying in the event venue is denoted by $\delta$. The schedule of the exposure risk must be incorporated in the simulations. To account for the departures and arrivals and other assumptions stated above, the expression for $\beta$ is revised as

$$\tilde{\beta} = \left(\frac{R}{\tau} + \frac{R}{\tau N}\right)\delta.$$

All parameters and state variables, except $c$, are nonnegative. The $\alpha$ is multiplied by $\tau$ to adjust the timescale since the number of days for surveillance is 14 days as recommended by the *World Health Organization (2020c)*. The model is represented by

$$\frac{dS}{dt} = -\tilde{\beta} S \frac{I}{N} + \alpha E + c$$

$$\frac{dE}{dt} = \tilde{\beta} S \frac{I}{N} - (1-\alpha)\tau E - \alpha\tau E = \tilde{\beta} S \frac{I}{N} - \tau E$$

$$\frac{dI}{dt} = (1-\alpha)\tau E$$

where $N = S + E + I(0) + 1$ (the constant 1 assures presence of at least one person in the venue of the event) and $N$ is less than or equal the maximum capacity of the event venue. The initial condition for $I$ is assumed to be one infected. In the Philippines, there are few confirmed cases but more than 200 persons under investigation as of 07 February 2020 (*Modesto 2020*). The probability of having this one initial case could be low in a country where the disease is possibly contained; but to account for the uncertainty, it is assumed that there is one possible case.

Insights that can be drawn from this model are as follow:

Insight A. The exposure time (number of hours an infectious person is staying in the event venue) is a significant factor in spreading the disease (Figure 2). Under the assumptions used, an infected person staying more than 9 hours could infect other people.

Insight B. The level of risk asymptotically increases when population size of the susceptible increases but the exposure time dictates if disease transmission will occur (Figure 3 and 4).

Insight C. Assuming the exposure time is 18 hours, an exposed person could be protected from being infected if the level of protection is >70 percent (Figure 5).

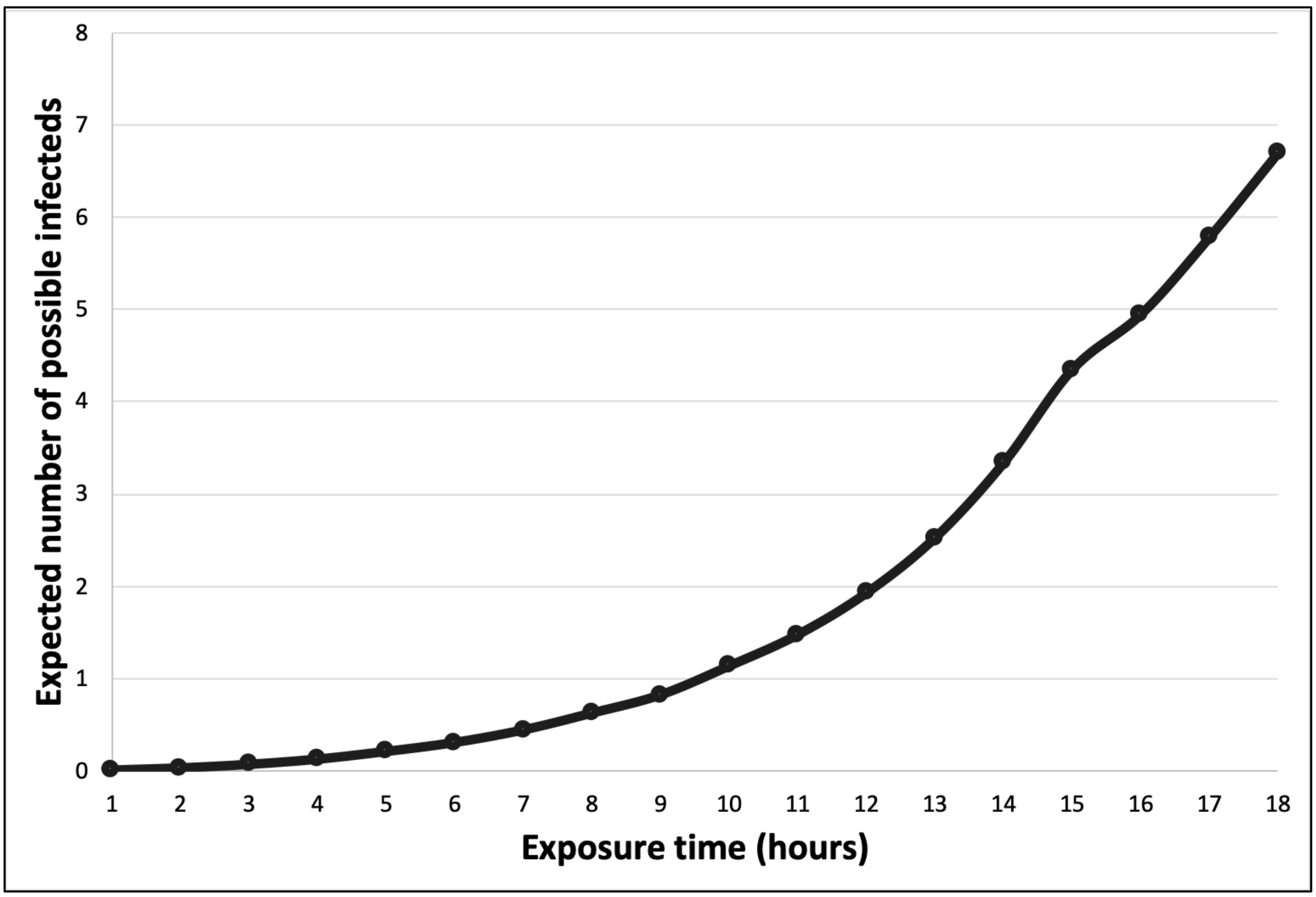

**Figure 2**. Effect of exposure time. An infected person staying more than 9 hours could infect other people (i.e., when expected number of possible new infecteds≥1). Parameters used: $\alpha = 0$, $c = 0$, $S_0 = 10{,}000$. The y-axis counts the possible new cases infected by the primary case

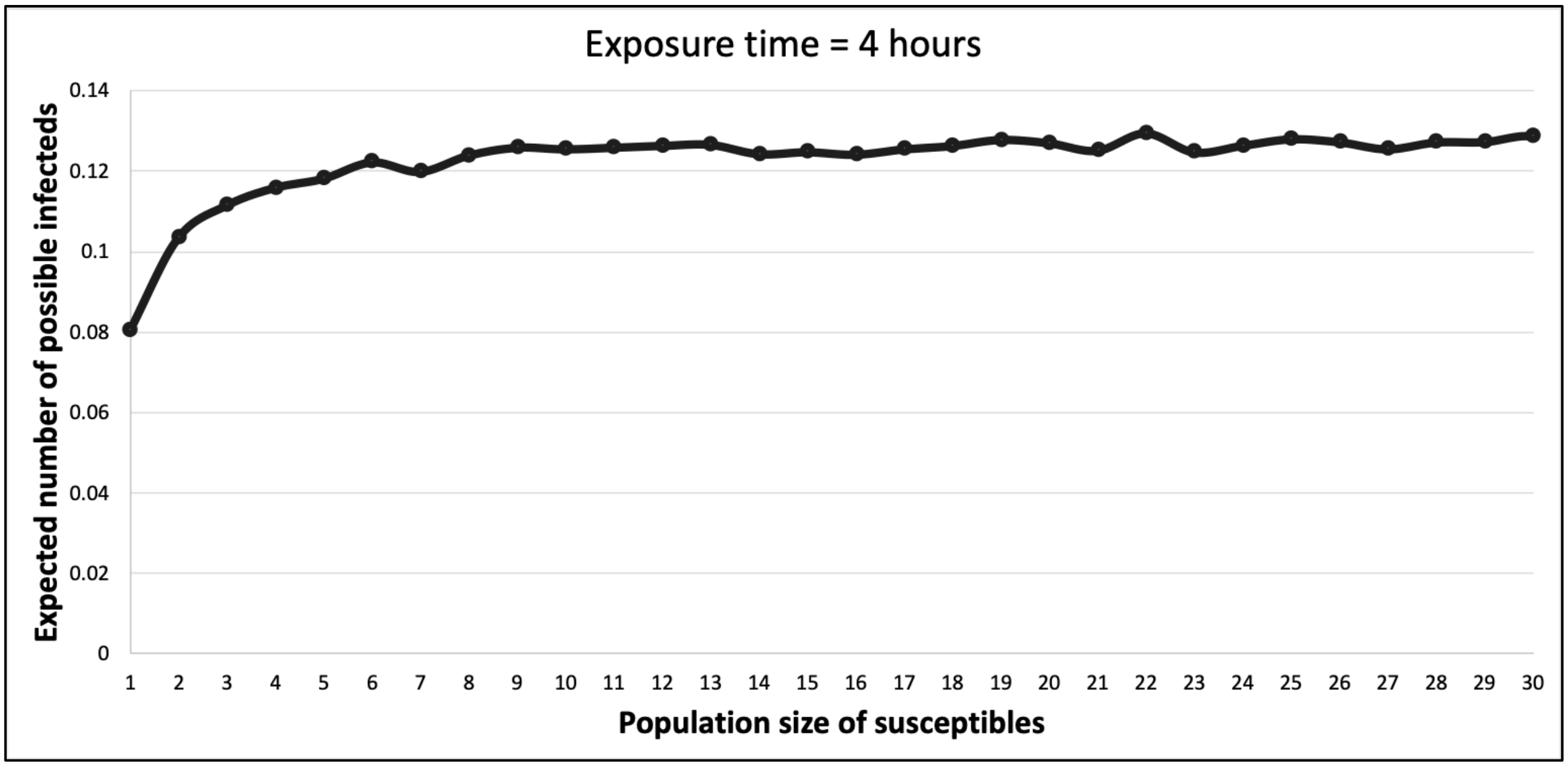

**Figure 3**. Risk of disease transmission increases as population size of susceptibles increases, $\delta = 4$. Having an exposure time = 4 hours, there is a low chance of disease transmission. Parameters used: $\alpha = 0$, $c = 0$. The y-axis counts the possible new cases infected by the primary case

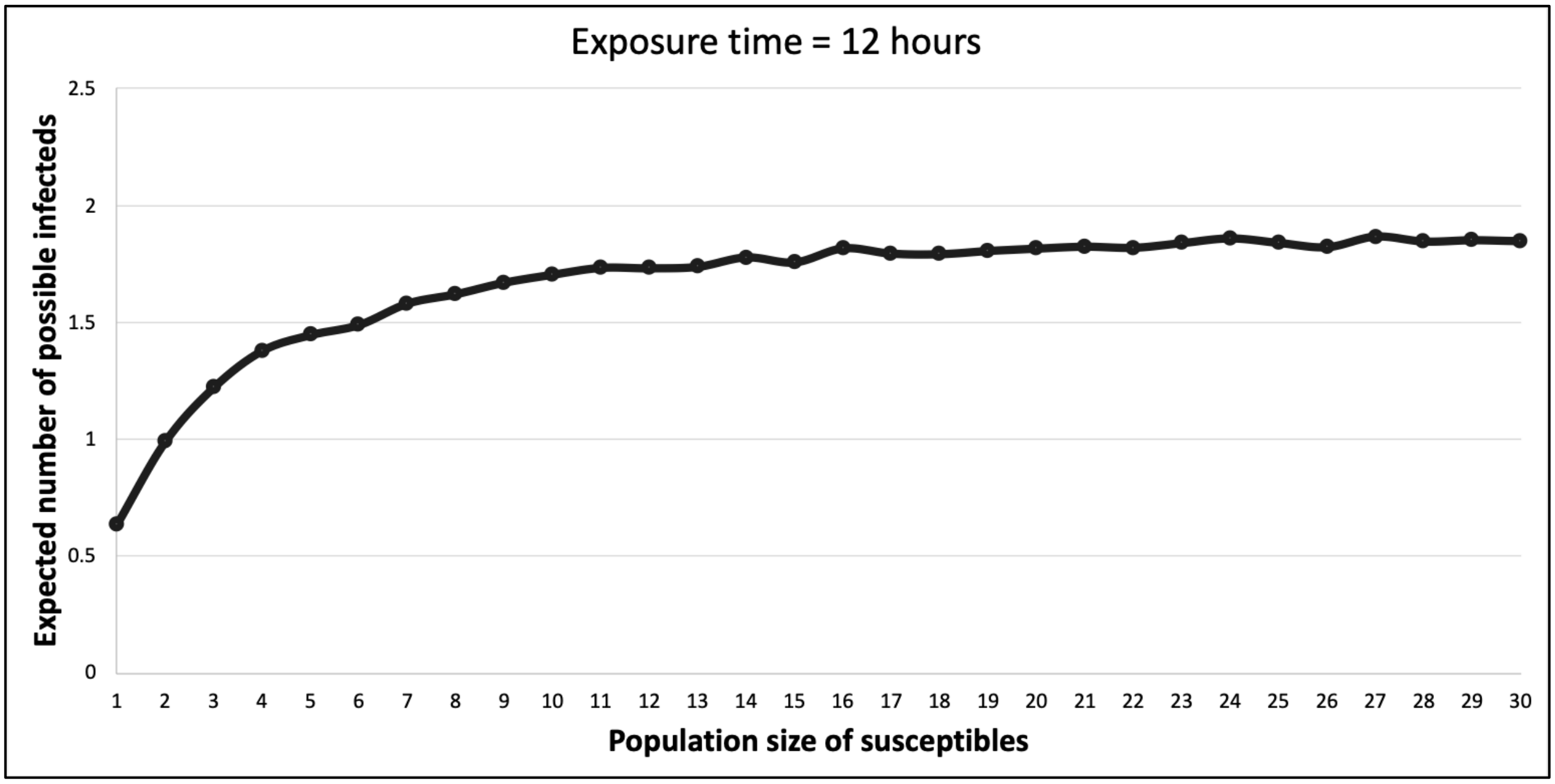

**Figure 4**. Risk of disease transmission increases as population size of susceptibles increases, $\delta = 12$. Having an exposure time = 12 hours, new cases may occur in an event with more than 2 attendees. Parameters used: $\alpha = 0$, $c = 0$. The y-axis counts the possible new cases infected by the primary case

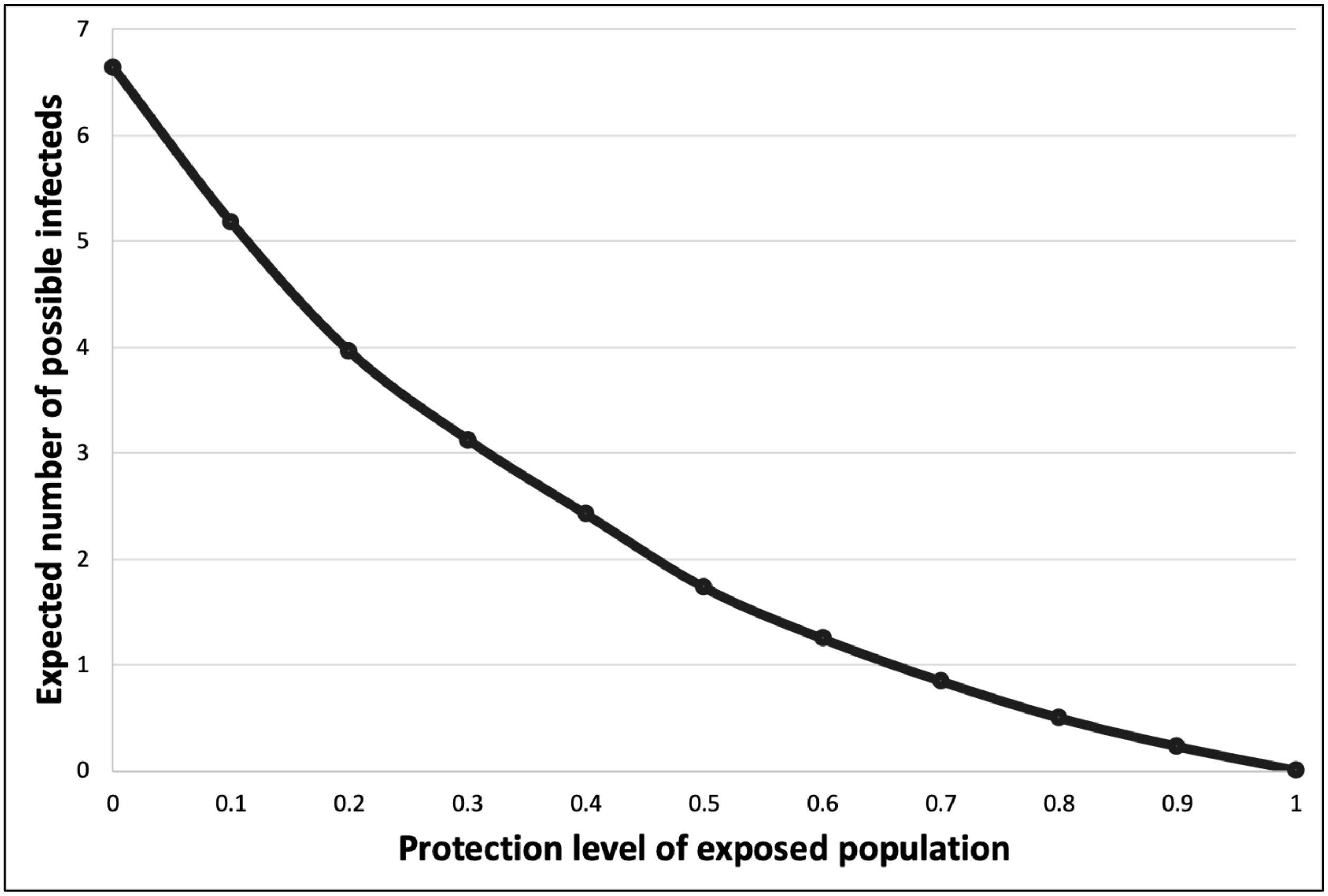

**Figure 5**. Significance of having a protection. An exposed person could be protected from being infected if the level of protection is >70%. Parameters used: $\delta = 18$, $c = 0$, $S_0 = 10{,}000$. The y-axis counts the possible new cases infected by the primary case

## CONCLUSIONS AND RECOMMENDATIONS

Results of the mathematical models agree with the recommendation of *World Health Organization (2020b)* that "it is still possible to interrupt virus spread, provided that countries put in place strong measures to detect disease early" – e.g., development of rapid diagnostic tests, and increasing effectiveness of passenger screening in airports in which thermal screening for 2019-nCoV infection is estimated using simulation to be 46 percent (95% CI 36.0-58.0) ineffective (*Quilty et* *al. 2020*), – “isolate and treat cases, trace contacts, and promote social distancing measures commensurate with the risk.” Individuals can also practice basic protective measures, especially in high risk areas (*World Health Organization 2020d)*. In the Philippines, many big events are cancelled because of the risk (*Malasig 2020*). In Singapore, the government raised its Disease Outbreak Response System Condition (DORSCON) level to Orange (*Mohan and Baker 2020*), which means there will be moderate disruption on daily life because “disease is severe and spreads easily from person to person, but the disease has not spread widely in Singapore and is being contained.”

The author formulated a model that can be used to recommend auxiliary strategies to prevent the spread of 2019-nCoV ARD in large social gatherings, especially during the early period of a possible outbreak. The model shows that the exposure time is a significant factor in spreading the disease. An infected person staying more than 9 hours could infect other people. Assuming the exposure time is 18 hours, the model recommends that attendees of the social gathering should have a protection with more than 70 percent effectiveness.

Users of the models should be critical in reviewing the model structure, assumptions and results before applying the insights drawn to decision making. Preliminary models can be too simplistic due to limited information available (*Cohen 2020b*), and do not consider detailed effect of the heterogeneity in susceptible and patient profiles. There can be infecteds who are super-spreaders or super-virus-shedders, and there can be infecteds who are isolated. Virus transmission in different settings or locations may also vary, e.g., transmission inside households, hospitals, offices, cruise ships and airports may have higher transmission rates compared to an open space such as park. A continuously updated global risk assessment model, e.g., using network science (*Robert Koch Institut 2020*), can be useful in visualization and drawing 'intuition'. Socio-economic and political factors (e.g., long duration of lockdown and travel ban) as well as influence of media can also affect the dynamics of the epidemics. In the midst of uncertainties, an effective risk communication and community engagement (RCCE) is essential to prevent infodemics, an excessive amount of information including disinformation and misinformation, that can be detrimental to public health control strategies. Moreover, the virus itself can mutate as it circulates from human to human, and other routes of transmission maybe possible (e.g., fecal-oral transmission, aerosol transmission). As the situation continuously progresses, continuous surveillance and updated predictions are necessary, hence, prescriptions of models may change.

**Acknowledgement:**

JFR is supported by The Abdus Salam International Centre for Theoretical Physics Associate Scheme, Trieste Italy.

**Declarations:**

The author declares that there are no conflicts of interest.

**APPENDIX**

Additional Simulations:

Suppose

- $\tau = Normal(14{,}25) \times 24$ where $Normal(14{,}25)$ characterizes normally distributed random numbers with mean = 14 and variance = 25
- $R = 2 + Poisson(2)$
- $c = 0, S_0 = 1{,}000$
- Mean denotes the average of 1000 simulation runs; Stdev is the standard deviation

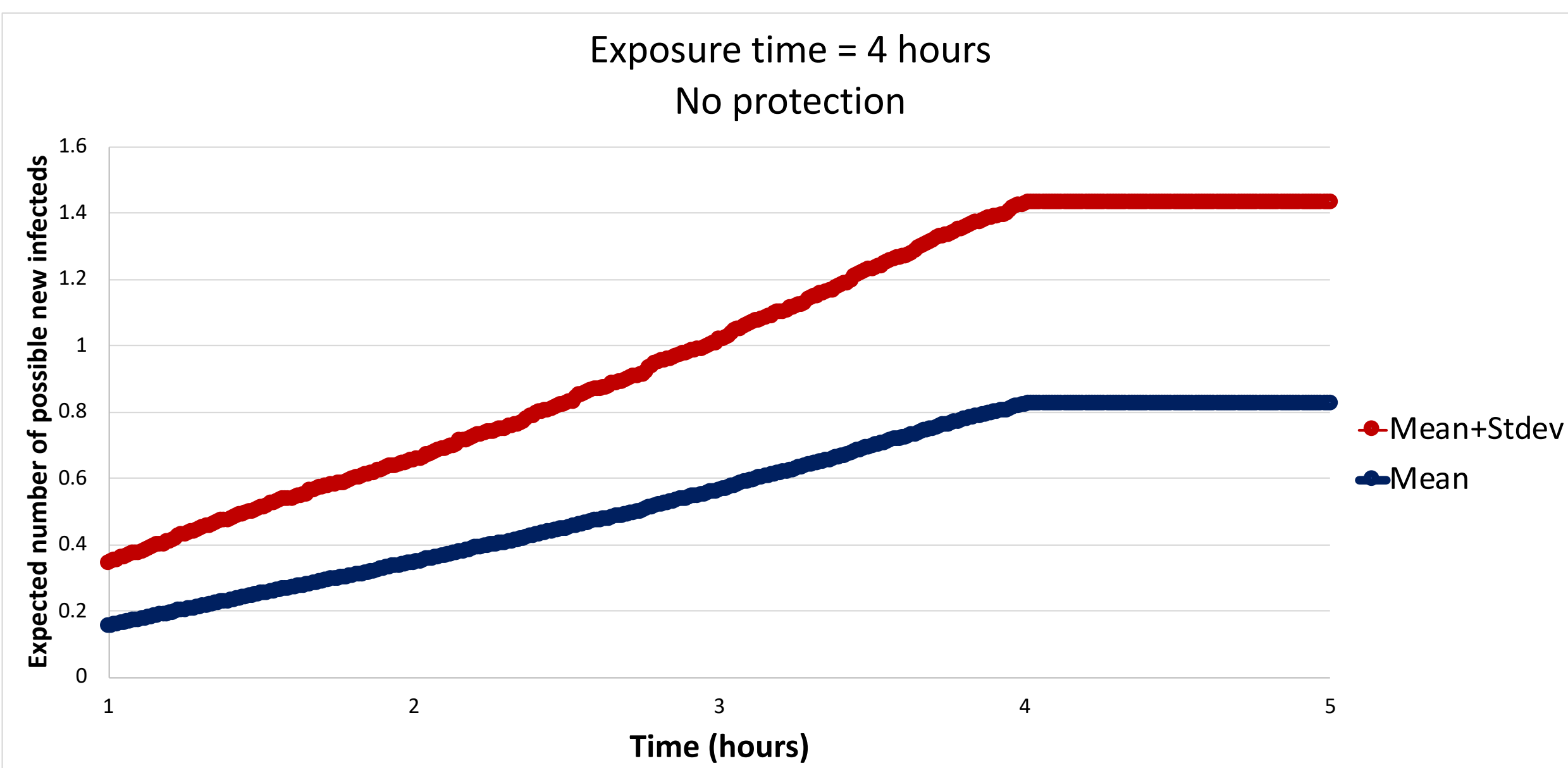

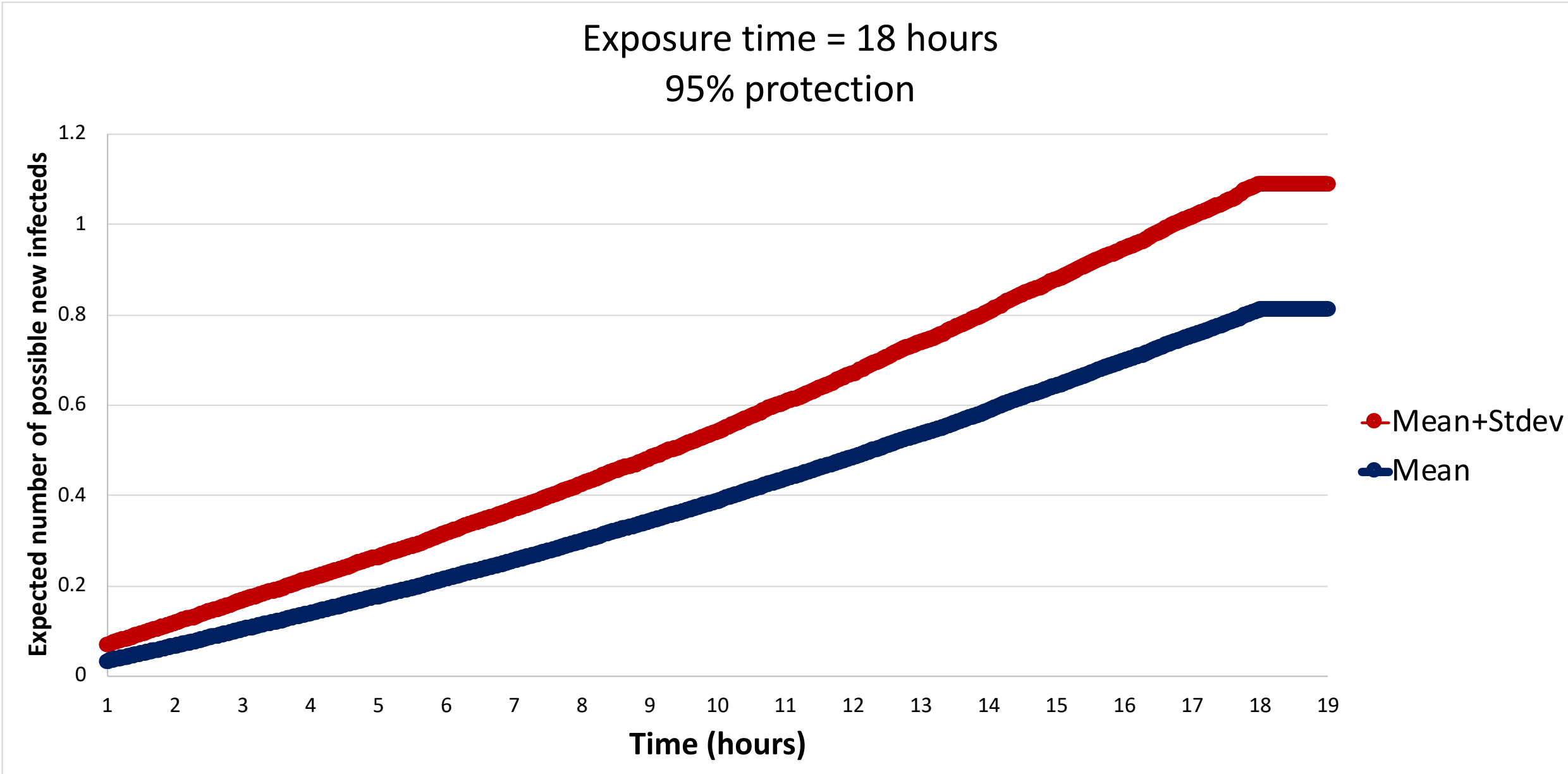
Exposure time = 18 hours
95% protection
Expected number of possible new infecteds
1.2
1
0.8
0.6
0.4
0.2
0
1 2 3 4 5 6 7 8 9 10 11 12 13 14 15 16 17 18 19
Time (hours)
Mean+Stdev
Mean